\documentclass[conference,10pt,letterpaper]{IEEEtran}

\usepackage[utf8]{inputenc}
\usepackage{amsmath, amssymb}
\usepackage{mathtools}
\usepackage{siunitx}
\usepackage{bbm}
\usepackage{subcaption}
\usepackage{graphicx}
\usepackage{placeins}
\usepackage[linesnumbered, ruled, vlined]{algorithm2e}
\usepackage{epstopdf}
\usepackage[rmargin=0.62in, lmargin=0.62in, bmargin=1.02in, tmargin=0.7in]{geometry}
\usepackage{setspace}

\usepackage{color}
\usepackage{soul}
\usepackage[backend=bibtex, style=ieee]{biblatex}

\usepackage{tikz}
\usepackage{textcomp}
\usepackage{hyperref}
\usepackage{lipsum}

\newcommand\copyrighttext{%
  \footnotesize \textcopyright 20XX IEEE.  Personal use of this material is permitted.  Permission from IEEE must be obtained for all other uses, in any current or future media, including reprinting/republishing this material for advertising or promotional purposes, creating new collective works, for resale or redistribution to servers or lists, or reuse of any copyrighted component of this work in other works.}
\newcommand\copyrightnotice{%
\begin{tikzpicture}[remember picture,overlay]
\node[anchor=south,yshift=10pt] at (current page.south) {\fbox{\parbox{\dimexpr\textwidth-\fboxsep-\fboxrule\relax}{\copyrighttext}}};
\end{tikzpicture}%
}
\title{QoS-Aware Scheduling in New Radio\\Using Deep Reinforcement Learning}

\definecolor{airforceblue}{rgb}{0.36, 0.54, 0.66}

\SetKw{Continue}{continue}
\SetKw{Break}{break}

\makeatletter
\newcommand\RedeclareMathOperator{%
  \@ifstar{\def\rmo@s{m}\rmo@redeclare}{\def\rmo@s{o}\rmo@redeclare}%
}
\newcommand\rmo@redeclare[2]{%
  \begingroup \escapechar\m@ne\xdef\@gtempa{{\string#1}}\endgroup
  \expandafter\@ifundefined\@gtempa
     {\@latex@error{\noexpand#1undefined}\@ehc}%
     \relax
  \expandafter\rmo@declmathop\rmo@s{#1}{#2}}
\newcommand\rmo@declmathop[3]{%
  \DeclareRobustCommand{#2}{\qopname\newmcodes@#1{#3}}%
}
\@onlypreamble\RedeclareMathOperator
\makeatother

\DeclareRobustCommand*{\IEEEauthorrefmark}[1]{%
  \raisebox{0pt}[0pt][0pt]{\textsuperscript{\footnotesize #1}}%
}

\renewcommand{\vec}[1]{\boldsymbol{#1}}
\DeclareMathOperator{\Ex}{\mathbb{E}}
\RedeclareMathOperator{\Pr}{\mathbb{P}}

\DeclareMathOperator{\Indicator}{\mathbbm{1}}
\DeclareMathOperator{\grad}{\nabla}

\author{
\IEEEauthorblockN{Jakob Stigenberg\IEEEauthorrefmark{1},  Vidit Saxena\IEEEauthorrefmark{2}, Soma Tayamon\IEEEauthorrefmark{2}, Euhanna Ghadimi\IEEEauthorrefmark{2}}
\vspace{0.1in}
\IEEEauthorblockA{\IEEEauthorrefmark{1}stigen@kth.se}
\IEEEauthorblockA{\IEEEauthorrefmark{2}Ericsson AB, Stockholm, Sweden
\\\{vidit.saxena, soma.tayamon, euhanna.ghadimi\}@ericsson.com}
}

\date{\today}

\begin{document}

\setlength{\textfloatsep}{0.1cm}
\setlength{\floatsep}{0.1cm}

\maketitle
\copyrightnotice

\begin{abstract}
Fifth-generation (5G) New Radio (NR) cellular networks support a wide range of new services, many of which require an application-specific quality of service (QoS), e.g. in terms of a guaranteed minimum bit-rate or a maximum tolerable delay. Therefore, scheduling multiple parallel data flows, each serving a unique application instance, is bound to become an even more challenging task compared to the previous generations. Leveraging recent advances in deep reinforcement learning, in this paper, we propose a QoS-Aware Deep Reinforcement learning Agent (QADRA) scheduler for NR networks. In contrast to state-of-the-art scheduling heuristics, the QADRA scheduler explicitly optimizes for the QoS satisfaction rate while simultaneously maximizing the network performance. Moreover, we train our algorithm end-to-end on these objectives. We evaluate QADRA in a full scale, near-product, system level NR simulator and demonstrate a significant boost in network performance. In our particular evaluation scenario, the QADRA scheduler improves network throughput by $30\%$ while simultaneously maintaining the QoS satisfaction rate of VoIP users served by the network, compared to state-of-the-art baselines.
\end{abstract}

\section{Introduction}
Traffic in cellular networks has increased dramatically in the recent decades. The fifth generation (5G) mobile network, also known as New Radio (NR), is designed to further increase traffic capacity, provide support for new use-cases, and enhance development of a diverse set of applications, including internet of things (IoT) applications and autonomous and/or remotely controlled systems \cite{dahlman20185g}. With an increasingly diverse set of applications in NR networks, maintaining the quality of service (QoS), compared to previous generations of cellular networks, is bound to become an even more challenging task. The scheduler is one of the core components of cellular networks and controls the allocation of the finite set of network resources, i.e. time, frequency, and spatial resources, to user equipment (UE) in both uplink and downlink transmissions. The development of efficient schedulers is therefore key for efficient usage of network resources and to overall network performance \cite{capozzi2012downlink}.

\begin{figure}
    \centering
    \includegraphics[width=\linewidth]{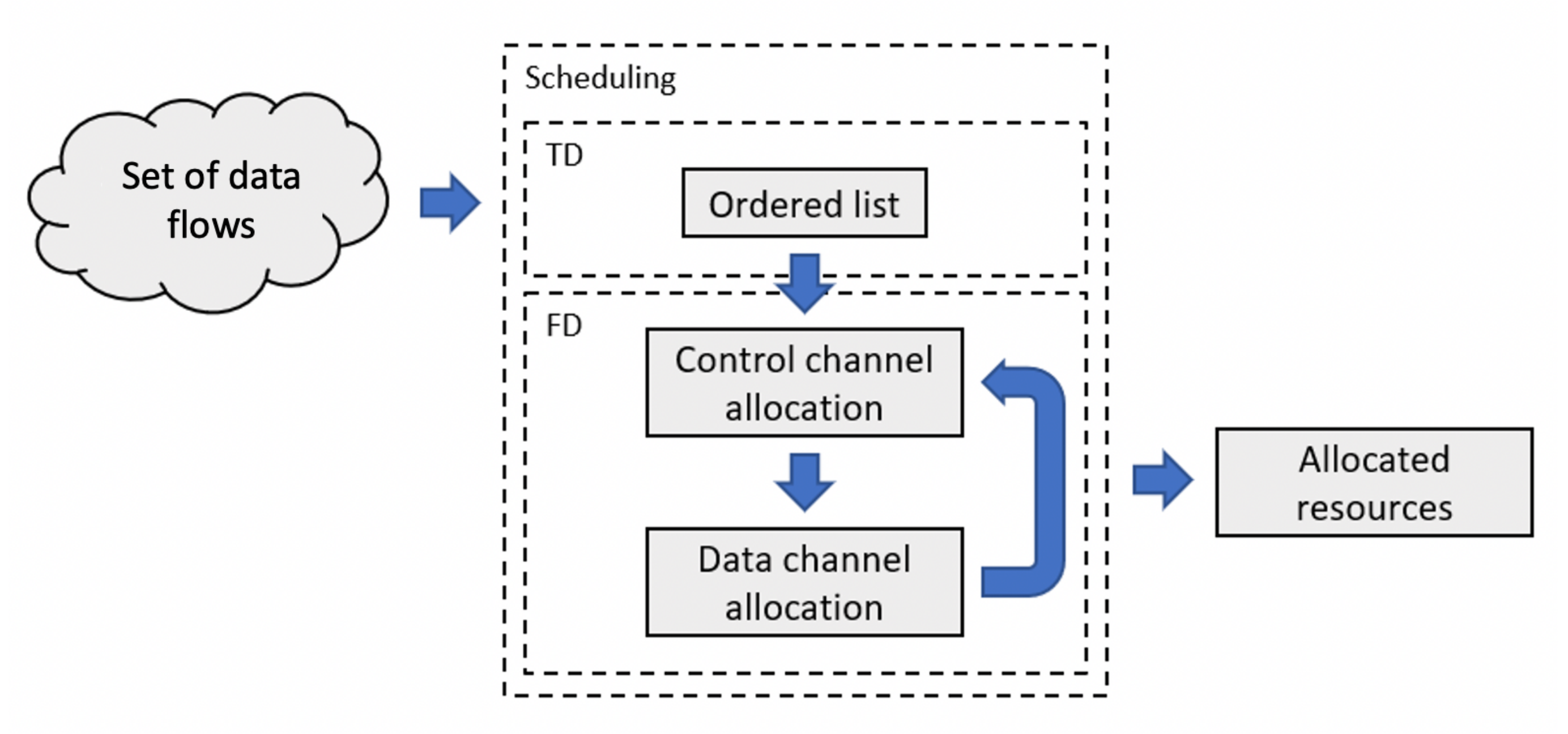}
    \caption{Overview of the scheduling process. Given a set of data flows, the TD scheduler sorts the flows in descending order of priority. The FD scheduler then allocates resources according to the priority assigned to each flow.}
    \label{fig:scheduling-overview}
\end{figure}

The scheduling algorithm is typically separated into two independent operational components, a time-domain (TD) unit, selecting the data flows to be served in each scheduling instance, and a frequency-domain (FD) unit, which allocates communication resources in the physical radio domain to the selected data flows \cite{capozzi2012downlink, monikandan2017review, grondalen2016scheduling}. This is shown in Fig. \ref{fig:scheduling-overview} and will be further detailed in Sec. \ref{sec:model-objectives}. 

Current state-of-the-art TD scheduling algorithms are based on heuristics that assign numerical weights to data flows based on properties such as channel quality, head-of-line delay and average throughput. A data flow is then given priority based on its weight, relative to other data flows \cite{capozzi2012downlink, monikandan2017review, grondalen2016scheduling}. Data flows are thus \emph{sorted}, implicitly, based on the assigned weights. However, sorting data flows based on heuristics defined through individual data flows may ultimately result in a sub-optimal scheduling strategy. Firstly, by failing to account for the composition of data flows, this approach is unable to consider the constraints enforced on the FD unit, i.e. physical resource constraints. Secondly, heuristic approaches are unable to explicitly optimize for network performance indicators, e.g. QoS satisfaction rate and network throughput.

Moreover, with the increased traffic diversity in NR networks as well as the increased flexibility offered by the NR specification \cite{dahlman20185g}, reliably maintaining and tuning network performance through heuristics might become a complex and daunting task.

To overcome these issues, in this paper, we propose a new approach to TD scheduling for NR using recent advancements in deep reinforcement learning (DRL) \cite{mnih2013playing}. Using DRL, we formulate, train, and evaluate a TD scheduling algorithm that
\begin{enumerate}
    \item \emph{explicitly} sorts data flows by taking all data flows into account as well as their QoS requirements,
    \item explicitly optimizes the aggregate network performance as well as QoS satisfaction rate,
    \item is trained end-to-end on the optimization objective, and
    \item allows for fine-granular control of fundamental network performance trade-offs by configuring the training objective with a, so-called, preference vector.
\end{enumerate}
We denote our approach as QoS-Aware Deep Reinforcement learning Agent (QADRA) for NR Time-domain Scheduling.

The remainder of this article is structured as follows. First, in Section \ref{sec:model-objectives}, we discuss the scheduling model used in which we develop our algorithm. In Section \ref{sec:drl}, we provide a brief overview of DRL and in Section \ref{sec:method} we present our main contribution, the QADRA scheduler. Finally, we evaluate the approach in Section \ref{sec:exp-res} using a full-scale NR system level simulator, and conclude the paper in Section \ref{sec:conclusion}.


\section{Model and objectives} \label{sec:model-objectives}

Consider a TD-FD scheduler tasked with assigning physical radio resources to data flows \cite{capozzi2012downlink, grondalen2016scheduling, monikandan2017review}. We denote a data flow with $\vec{f}\coloneqq\{\vec{x},\vec{q}\}$, where $\vec{x}\in\mathcal{R}$ denotes the data flow feature vector in some feature space $\cal R$ and $\vec{q}\in\mathcal{Q}$ denotes the flow QoS requirements in some space of QoS requirements, $\mathcal{Q}$. Data flows that lack strict QoS requirement, such as file transfers, are referred to as \emph{best effort} traffic. At every transmission time interval (TTI) indexed by $t=1,2,\dots$, the scheduling process is executed in the following manner: the scheduler first  receives an input set of $N_t$ data flow feature vectors $\mathcal{X}_t\coloneqq\{\vec{x}_{t,j}\}_{j=1}^{N_t}$. The TD scheduler compiles an output list $\mathcal{Y}_t$ of data flows by sorting the elements of $\mathcal{X}_t$ in a decreasing order of priority inferred from their respective feature vectors. Next, the FD scheduler parses $\mathcal{Y}_t$ serially and assigns physical radio resources to each data flow until the available resources are exhausted. The scheduled data flows are then transmitted over their corresponding physical resource assignments. This is summarized in Fig. \ref{fig:scheduling-overview}.

Physical radio resources carry the control channel as well as the data channel. The control channel contains the information required to locate and decode the data payload, carried by the data channel, at the receiver. In state-of-the-art cellular networks, the physical resources are divided between the physical uplink and downlink control channels (PUCCH and PDCCH, respectively), and the physical uplink and downlink shared channels (PUSCH and PDSCH, respectively). The typical FD scheduling process hence consists of two sequential steps: control resource allocation and data resource allocation, as seen in Fig. \ref{fig:scheduling-overview}. With control resource allocation, the FD scheduler assigns PDCCH and PUCCH resources to the highest prioritized data flows, as defined by the TD scheduler. The FD scheduler then tries to allocate data channel resources, PUSCH or PDSCH, to these data flows. Starting with the most prioritized data flow, data channel resources are assigned if available, otherwise, its assigned control channel resources are deleted. The process is then repeated for each data flow until all data flows with control channel allocation have been attempted to be given data channel resources. 

The delay experience by a data packet, which is served by a data flow, is defined as the time between the creation and the successful reception of the packet by the intended receiver. Further, the average network throughput is defined in terms of the number of bits successfully transmitted over the network per time unit.



\section{Deep reinforcement learning} \label{sec:drl}
Reinforcement learning (RL) \cite{sutton2018reinforcement} is a paradigm of artificial intelligence that deals with learning optimal (sequential) decision making in dynamic environments. It is based on Markov Decision Processes (MDPs), characterized by a state space $\cal S$, action space $\cal A$, state transition function $T(s', s, a) = \Pr\big[S_{t+1}=s' \big| S_t=s, A_t = a\big]$, reward function $R(s', s, a)$, and discount factor $\gamma\in[0,1)$. The general RL algorithm aims to learn a policy distribution
\begin{equation}
    \pi: \mathcal{S}\times\mathcal{A} \to [0, 1]
\end{equation}
that maximizes the (expected and discounted) accumulated reward:
\begin{equation}
    \pi^* = \arg\max_\pi \Ex_\pi\Big[\sum\nolimits_{t=0}^\infty \gamma^t R(S_{t+1}, S_t, A_t) \Big| S_0 = s_0\Big],
\end{equation}
where $A_t \sim \pi(S_t, \cdot)$ and $S_{t+1}\sim T(\cdot, S_t, A_t)$. Note, the transition function, $T$, and reward function, $R$, are unknown a priori, thus requiring an algorithm to explore the environment dynamics in order to learn the optimal action for an arbitrary state.

\subsection{Deep Q-Networks}
Deep Q-Networks (DQNs) \cite{mnih2013playing} were one of the first approaches to incorporate deep neural networks into an RL algorithm. DQNs are based on Q-learning \cite{watkins1989learning}, and use deep neural networks to learn the value of executing a certain action, $a$, in a given state, $s$, $Q(s, a)$. In particular, the Q-function, $Q: \mathcal{S}\times\mathcal{A} \to \mathbb{R}$, denotes the (expected and discounted) accumulated reward following a specific state-action pair under an optimal policy,
\begin{equation}
    Q(s, a) = \Ex_{\pi^*}\Big[\sum\nolimits_{t=0}^\infty \gamma^t R(S_{t+1}, S_t, A_t) \Big| S_0 = s, A_0 = a \Big],
\end{equation}
with $S_{t+1}\sim T(\cdot, S_t, A_t)$ and $A_t\sim\pi^*(S_t, \cdot)$. Given the Q-function, an optimal policy is given by
\begin{equation}
    \pi^*(s, a) = \Indicator\{a = a^*\},
\end{equation}
where $a^* = \arg\max_{a\in\mathcal{A}} Q(s, a)$ and where $\Indicator\{\cdot\}$ denotes the indicator function.

\section{Time domain scheduling using Deep Reinforcement Learning} \label{sec:method}
\begin{figure}
    \centering
    \includegraphics[width=\linewidth]{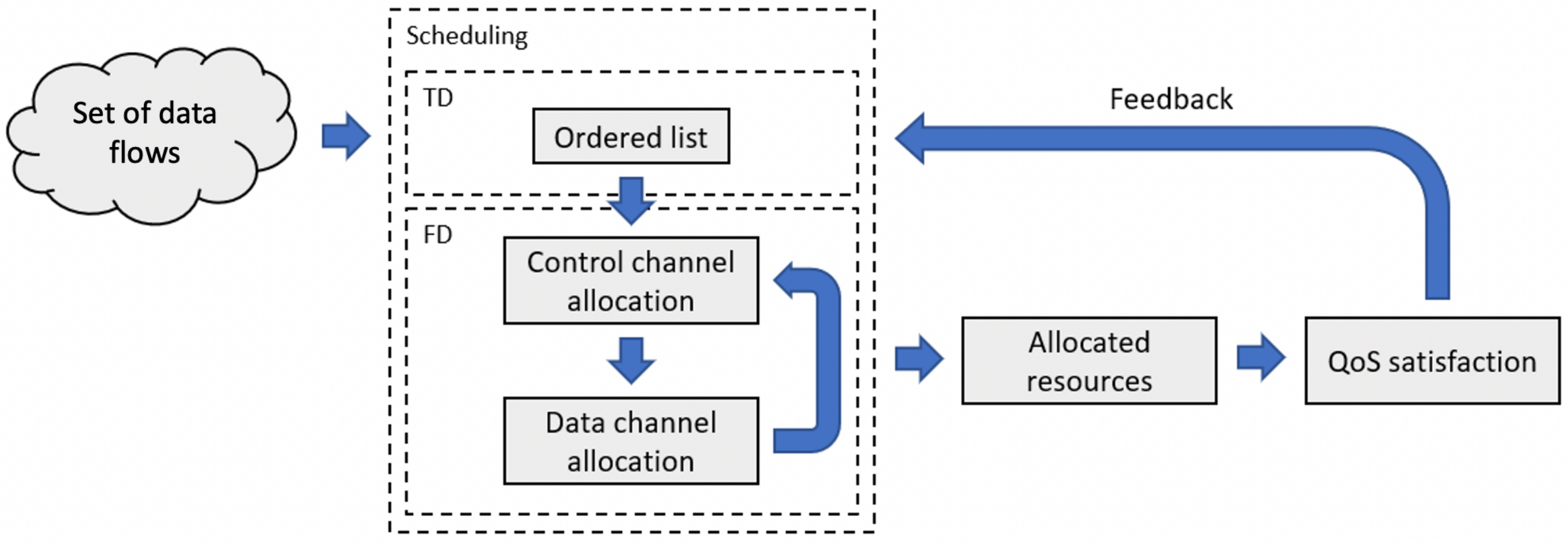}
    \caption{Block diagram of QoS-aware scheduling algorithm proposed in this paper. The TD scheduling unit is trained with the output of the FD scheduling unit made available through a feedback loop.}
    \label{fig:scheduling-feedbackloop}
\end{figure}

In this section we present the QADRA scheduling algorithm, a new approach to TD scheduling in NR using DRL. At a high level, QADRA is a machine learning-guided list sorting algorithm, which is trained using the feedback loop shown in Fig. \ref{fig:scheduling-feedbackloop}.

Given the scheduling process discussed in Sec. \ref{sec:model-objectives}, the TD scheduling problem reduces to sorting of data flows. Current scheduling algorithms approach this problem by assigning weights to each data flow using heuristics \cite{capozzi2012downlink, grondalen2016scheduling, monikandan2017review}, and subsequently sort the data flows, implicitly, based on their assigned weights. We wish to construct an algorithm that learns to \emph{explicitly} sort data flows such that the final resource allocation is optimal in terms of the user QoS satisfaction rate.

We first define, through the reward function, the optimization objective of our algorithm. Avoiding any form of reward shaping, we define the reward based on the fulfillment of flow QoS requirements, encoded by $\vec{q}_{t,j}$, for flow $j$ during TTI $t$ as
\begin{equation}
    r_{t,j} = \begin{cases} 
0\quad\qquad \vec{q}_{t,j}~\text{satisfied},\\
-1\qquad\,\vec{q}_{t,j}~\text{not satisfied}
\end{cases}
.
\end{equation}
Best effort traffic, on the other hand, is given a reward equal to the number of bits transferred successfully. Next, we form a reward vector $\vec{r}_t = [r_{t,G_1},\ldots,r_{t,G_{N_G}}]^T$, where $G_i,\,i\in\{1,\dots,N_G\}$ denotes a data flow \emph{type} (i.e., a group of flows with identical QoS requirements, e.g. conversational video), $N_G$ denotes the total number of data flow types, and
\begin{equation}
    r_{t,G_i} = \sum_{j=1}^{N_t} \Indicator\{\vec{f}_{t,j}~\text{is of type}~G_i\} r_{t,j},
\end{equation}
the total reward gathered by all flows of type $G_i$ during TTI $t$. Finally, in order to reduce the vectorized reward, $\vec{r}_t$, into a scalar reward, $r_t$, we introduce a preference vector, $\vec{\omega}\in\mathbb{R}^{N_G}$, and let
\begin{equation}
    r_t = \vec{\omega}^T\vec{r}_t.
\end{equation}
The preference vector, thus, provides a way to explicitly control the trade-off between conflicting QoS requirements by modifying the optimization objective. For example, letting ${\omega_j=\Indicator\{i=j\}}$ results in pure optimization of the QoS of data flows in $G_i$.

\subsection{List sorting using deep reinforcement learning} \label{sec:method:sorting}
\begin{algorithm}
\SetAlgoLined
\KwData{A list $\mathcal{X} = \{x_j\}_{j=1}^N$}
\KwResult{A reordered list, $\mathcal{Y}$}
Initialize $\mathcal{Y}=\emptyset$\;
\For{$k=0,1,\ldots,N-1$}{
    Select $a_k\in\{1,\ldots,N-k\}$\; \label{alg:selection-sort:policy}
    Append $x_{a_k}$ to $\mathcal{Y}$\;
    Remove $x_{a_k}$ from $\mathcal{X}$\;
}
\caption{Selection sort}
\label{alg:selection-sort}
\end{algorithm}
A key difficulty in applying RL to list sorting is the size of the action space. Given a set of $N$ unique elements, $x_1, \ldots, x_N$, there are $N!$ possible unique sequences, hence an action space of size $N!$. We circumvent this problem by reducing the sorting problem from one action, in a space of size $N!$, into a sequence of actions in much smaller spaces. In particular, the selection sort algorithm, shown in Algorithm \ref{alg:selection-sort}, sorts a list of $N$ unique elements in $N$ steps. At each step $k=0,1,2,\ldots,N-1$, the input list, $\cal X$, contains $N-k$ elements from which a policy selects one to append to the output list, $\cal Y$. Thus, using selection sort, $N$ actions in action spaces of sizes $N, N-1,\ldots, 1$ are sufficient.

The selection of $a_k\in\{0,1,\ldots,N-k\}$ (line \ref{alg:selection-sort:policy} of Algorithm \ref{alg:selection-sort}) fully determines the output order. For example, when sorting numerical lists, $x_j\in\mathbb{R}~\forall j$, letting $a_k=\arg\min_j x_j$ results in an output list in ascending order. In similar fashion, $N_t$ data flows can be sorted using a policy trained, e.g., using DRL. The policy is given the current input and output lists (of flow feature vectors) and returns the index of the data flow of the input list next to append to the output list. In other words, the state space is given by $\mathcal{S} = \mathcal{S}_i\times\mathcal{S}_o$, where $\mathcal{S}_i$ and $\mathcal{S}_o$ denote the state spaces of the input and output lists, respectively. Since the two lists both contain data flows, $\mathcal{S}_i=\mathcal{S}_o$. Each data flow is characterized by a feature vector $\vec{x}\in\mathcal{R}$ and a list may contain any number of data flows, hence
\begin{equation}
    \mathcal{S}_i=\mathcal{S}_o = 0\cup\mathcal{R}\cup\mathcal{R}\times\mathcal{R}\cup\ldots,
\end{equation}
where $0$ denotes an empty list. The action space depends on the current input list. Given $N_t$ data flows to sort, the action space of the $k$th action is given by $\mathcal{A}_{t,k}=\{0,1,\ldots,N_t-k\}$.
\subsection{The QADRA Scheduler}
\begin{figure}
    \centering
    \includegraphics[width=0.8\linewidth]{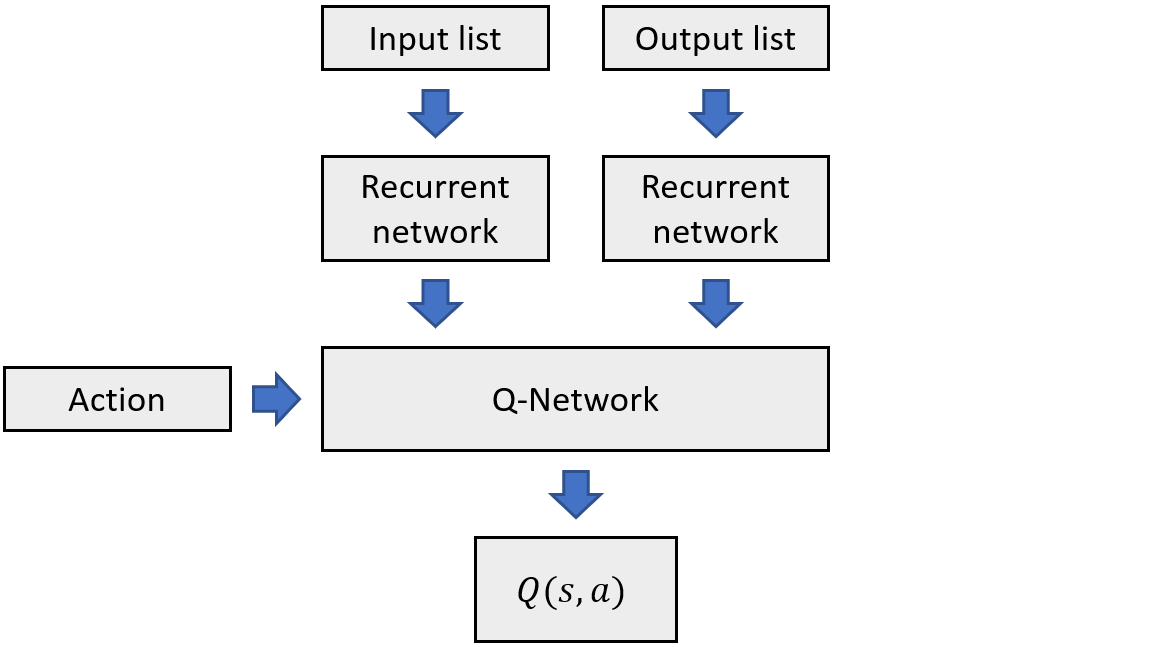}
    \caption{Overview of network architecture employed by the QADRA scheduler. Two recurrent networks capture the states of the input and output lists, $s$, that are then combined with an action, $a$, and evaluated using the Q-Network.}
    \label{fig:network-architecture}
\end{figure}

In this section, we describe the QADRA scheduler, which uses a selection sort approach to optimally schedule data flows in every TTI, in terms of the reward defined previously in this section. QADRA implements a, slightly modified, DQN as its selection policy. The states of the input and output lists are reduced to a state vector, $\vec{s}$, by sequentially passing each element of the respective lists through a recurrent neural network. A state-action pair is then evaluated by passing the state vector and the feature vector of the corresponding action through a feed-forward network, from here on denoted the Q-Network, as illustrated in Fig. \ref{fig:network-architecture}. Moreover, since sequential information is required to form the correct states during training, sequences of transitions are stored in the replay buffer used by the DQN algorithm, as opposed to individual transitions.


The QADRA scheduler consists of two distinct processes, allowing for distributed training, an actor process and a learner process. These are detailed below. During evaluation, only the actor process is run. We denote the recurrent networks using
\begin{equation}
    \vec{s}, \vec{d} \leftarrow h(\vec{x}, \vec{d}; \text{params}),
\end{equation}
where $\vec{d}$ represents the internal state of the network, $\vec{x}$ the input feature vector, and $\vec{s}$ denotes the output from the network. Two sets of parameters, $\phi, \theta$, are used, one for the input and one for the output list. The Q-Network is parameterized by $\psi$. When referring to all sets of parameters, we use $\Gamma = \{\phi, \theta, \psi\}$.

\subsection{Actor process}
The actor process is outlined in Algorithm \ref{alg:td-scheduling-algorithm:eval}. The actor explores the environment and stores sequences of states, actions and rewards into a replay buffer, then used by the learner process. The algorithm is given (and occasionally updated throughout training) sets of parameters $\phi, \theta, \psi$ as well as an exploration rate $\epsilon\in[0,1]$. During each TTI $t$, $N_t$ data flows are to be sorted whose feature vectors $\{\vec{x}_{t, k}\}_{k=1}^{N_t}$ form the input list $\mathcal{X}_t$. In order to save the initial set of data flows to the replay buffer, the input list is first cloned into $\mathcal{X}_t^-$. An output list, $\mathcal{Y}_t$, is initialized as well as internal states of the recurrent networks, $\vec{d}_i, \vec{d}_o$, and state vectors, $\vec{s}_i, \vec{s}_o$, with sub-scripts $i$ and $o$ referring to input and output lists, respectively. 

The input state vector, $\vec{s}_i$, is first calculated by sequentially passing each feature vector, $\vec{x}_{t, k}$, through the recurrent network parameterized by $\phi$. The input list is then sorted using $N_t$ actions, chosen using an $\epsilon$-greedy policy, as outlined in Section \ref{sec:method:sorting}. Following an action, the output state vector, $\vec{s}_o$, is updated accordingly. After $N_t$ actions, the output $\mathcal{Y}_t$ is given to the FD unit. A reward, $r_t$, is computed using the pre-defined preference vector, as previously discussed, and a new set of data flows, $\mathcal{X}_{t+1}$, are obtained. Finally, the observed sequence $(\mathcal{X}_t^-, a_{t,1},\ldots,a_{t,N_t}, r_t, \mathcal{X}_{t+1})$ is added to the replay buffer.

\IncMargin{0.01in}
\begin{algorithm}
\SetAlgoLined
\KwData{Recurrent network $h$ with two sets of params $\phi, \theta$. Feed-forward network $Q$ with params $\psi$. Exploration rate $\epsilon$.}
Let $\vec{s}$ denote the concatenation of $\vec{s}_i$ and $\vec{s}_o$\;
Acquire a set of data flows $\mathcal{X}_1=\{\vec{x}_{1,j}\}_{j=1}^{N_1}$\;
\For{$t=1,2,\ldots$}{
    Clone $\mathcal{X}_t \rightarrow \mathcal{X}_t^-$\;
    Initialize empty output list $\mathcal{Y}_t$\;
    Reset states: $\vec{s}_i,\vec{s}_o\leftarrow\vec{0}$\;
    Reset recurrent networks: $\vec{d}_i,\vec{d}_o\leftarrow\vec{0}$\;
    \For{$k=1,2,\ldots,N_t$}{
        $\vec{s}_i, \vec{d}_i \leftarrow h(\vec{x}_{t,k}, \vec{d}_i; \phi)$\;
    }
    \For{$k=1,2,\ldots,N_t$}{
        $a_{t,k} = \begin{cases}\arg\max_{j} Q(\vec{s}, \vec{x}_{t,j}; \psi) & \text{w.p.}~1-\epsilon \\ \sim\mathcal{U}(\{1,2,\ldots,N_t-k\}) & \text{w.p.}~\epsilon \end{cases}$\;
        $\vec{s}_o, \vec{d}_o \leftarrow h(\vec{x}_{t, a_{t,k}}, \vec{d}_o; \theta)$\;
        Append $\vec{x}_{t, a_{t, k}}$ to $\mathcal{Y}_t$ and remove from $\mathcal{X}_t$\;
    }
    Output $\mathcal{Y}_t$. Receive reward $r_t$ and new list $\mathcal{X}_{t+1}=\{\vec{x}_{t+1,j}\}_{j=1}^{N_{t+1}}$\;
    Add sequence $(\mathcal{X}_t^-, a_{t,1}, \ldots, a_{t, N_t}, r_t, \mathcal{X}_{t+1})$ to replay buffer\;
    Occasionally: update $\phi, \theta, \psi$ from the learner process\;
}
\caption{Actor process}
\label{alg:td-scheduling-algorithm:eval}
\end{algorithm}
\DecMargin{0.01in}

\subsection{Learner process}
\IncMargin{0.03in}
\begin{algorithm}
\SetAlgoLined
\KwData{Recurrent network $h$ with two sets of params $\phi, \theta$. Feed-forward network $Q$ with params $\psi$. Learning rate $\eta\in\mathbb{R}$. Let $\Gamma$ be short hand for $\{\phi, \theta, \psi\}$. Also, let $\vec{s}$ be the concatenation of $\vec{s}_i$ and $\vec{s}_o$}
Fixate $\phi^-, \theta^-, \psi^- \leftarrow \phi, \theta, \psi$\;
\While{training}{
    Sample $(\mathcal{X}_\tau, a_{\tau, 1}, \ldots,a_{\tau, N_\tau}, r_\tau, \mathcal{X}_{\tau+1})$ from the replay buffer\;
    
    Reset states: $\vec{s}_i,\vec{s}_i^-,\vec{s}_o,\vec{s}_o^-\leftarrow\vec{0}$\;
    Reset recurrent networks: $\vec{d}_i,\vec{d}_i^-,\vec{d}_o,\vec{d}_o^-\leftarrow\vec{0}$\;
    Reset gradient: $\grad\Gamma \leftarrow \vec{0}$\;
    
    \For{$k=1,2,\ldots,N_\tau$}{
        $\vec{s}_i, \vec{d}_i \leftarrow h(\vec{x}_{\tau, k}, \vec{d}_i; \phi)$\;
        $\vec{s}_i^-, \vec{d}_i^- \leftarrow h(\vec{x}_{\tau, k}, \vec{d}_i^-; \phi^-)$\;
    }
    
    \For{$k=1,2,\ldots,N_\tau-1$}{
        $\vec{s}_o^-, \vec{d}_o^- \leftarrow h(\vec{x}_{\tau, a_{\tau, k}}, \vec{d}_o^-; \theta^-)$\;
        $\grad \Gamma \leftarrow \grad \Gamma + \grad_\Gamma\frac{1}{2}\big[\gamma\max_j Q(\vec{s}^-, \vec{x}_{\tau, j}; \psi^-) - Q(\vec{s}, \vec{x}_{\tau, a_{\tau, k}}; \psi)\big]^2$\;
        $\vec{s}_o, \vec{d}_o \leftarrow h(\vec{x}_{\tau, a_{\tau, k}}, \vec{d}_o; \theta)$\;
        Remove $\vec{x}_{\tau, a_{\tau, k}}$ from $\mathcal{X}_\tau$\;
    }
    Reset target states: $\vec{s}_o^-, \vec{s}_i^- \leftarrow \vec{0}$\;
    Reset input target recurrent network: $\vec{d}_i^-\leftarrow\vec{0}$\;
    \For{$k=1,2,\ldots,N_{\tau+1}$}{
        $\vec{s}_i^-, \vec{d}_i^- \leftarrow h(\vec{x}_{\tau+1, k}, \vec{d}_i^-; \phi^-)$\;
    }
    $\grad\Gamma \leftarrow \frac{1}{N_\tau}\Big[\grad\Gamma + \grad_\Gamma\frac{1}{2}\big[r_\tau + \gamma\max_j Q(\vec{s}^-, \vec{x}_{\tau, j}; \psi^-) - Q(\vec{s}, \vec{x}_{\tau, a_{N_\tau}}; \psi)\big]^2\Big]$\;
    $\Gamma \leftarrow \Gamma - \eta\grad\Gamma$\;
    Occasionally: re-fixate $\phi^-, \theta^-, \psi^- \leftarrow \phi, \theta, \psi$\;
}
\caption{Learner process}
\label{alg:td-scheduling-algorithm:training}
\end{algorithm}
\DecMargin{0.03in}
The learner process is outlined in Algorithm \ref{alg:td-scheduling-algorithm:training}. The learner samples experiences, produced by the actor, from a replay buffer and updates the parameters of the networks to more accurately predict the value of state-action pairs, $Q(s, a)$. The algorithm is given a recurrent network, $h$, with two sets of parameters, $\phi, \theta$, and a Q-Network with corresponding parameters $\psi$. These networks are fixated (and occasionally re-fixated throughout training) with parameters $\phi^-, \theta^-, \psi^-$ and referred to as target networks (cf. \cite{mnih2013playing}). Each training step, an experience $(\mathcal{X}_\tau, a_{\tau,1},\ldots,a_{\tau, N_\tau},r_\tau,\mathcal{X}_{\tau + 1})$ is sampled from the replay buffer. The input and output states, $\vec{s}_i, \vec{s}_o$, and their target network equivalences, $\vec{s}_i^-, \vec{s}_o^-$, are initialized as well as the internal states of the recurrent networks, $\vec{d}_i, \vec{d}_i^-, \vec{d}_o, \vec{d}_o^-$. 

First, the input states, $\vec{s}_i, \vec{s}_i^-$ are computed by passing each data flow of $\mathcal{X}_\tau$ through the recurrent network, using parameters $\phi, \phi^-$, respectively. Then, at each step $k$, the target value must be computed, hence the target output state is first updated using the action $a_{\tau, k}$, followed by the accumulation of gradient, $\grad\Gamma$, on the squared error between the value of the chosen state-action pair and the value of the following state. Then, the output state is updated and prepared for the next action. The last action in each sequence is treated differently. A non-zero reward is given, hence included in the loss function and the next state is given by the next set of data flows to sort, $\mathcal{X}_{\tau + 1}$. Hence, the target input state must be recomputed, and the target output state reset. Following this step, the accumulated gradient is normalized by the sequence length, $N_\tau$, and a gradient descent step is performed.

\section{Experiments and results} \label{sec:exp-res}
\begin{table}[t]
    \centering
    \begin{tabular}{c|c}
         Parameter name & Value \\
         \hline
         no. of cells & 1 \\
         Cell radius & 166.67m \\
         UE mobility & Fixed \\
         Transmission scheme & FDD\\
         Carrier frequency & 0.6GHz \\
         Bandwidth & 10MHz \\
         Sub-carrier spacing & 30kHz \\
         no. of BS antennas & 4 \\
         no. of UE antennas & 4 \\
    \hline
    \end{tabular}
    \caption{Summary of simulation parameters.}
    \label{tab:sim-parameters}
   \end{table}
We train and evaluate our proposed algorithm, the QADRA scheduler, in a full scale NR system level simulator. Network parameters used in all runs are shown in Table  \ref{tab:sim-parameters}.
\subsection{Experimental setup}

\begin{figure}
    \centering
    \includegraphics[width=0.8\linewidth]{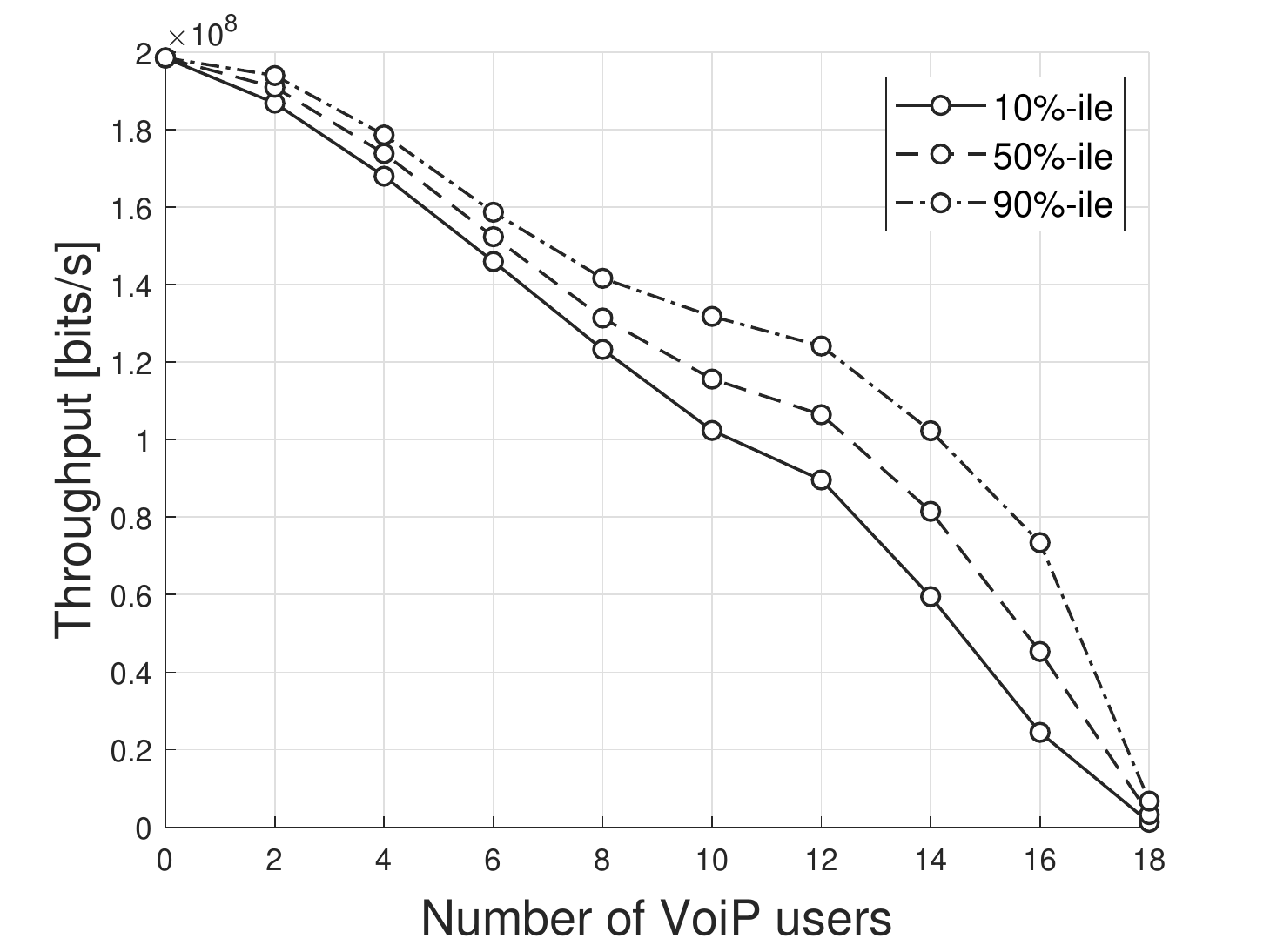}
    \caption{Starvation in practice. One full buffer downlink data flow and a varying number of VoIP users, each served by two data flows. The downlink cell throughput is greatly reduced with increasing number of VoIP data flows.}
    \label{fig:starvation}
\end{figure}

We deploy our algorithm in a scenario with the following traffic model: a single full buffer downlink operating simultaneously with ten VoIP users, each of which comprises an uplink and a downlink data flow. All UE have relatively good channel conditions (see Table \ref{tab:sim-parameters}). The particular scenario is prone to resource starvation (cf. \cite{pedroso2019low}) with increasing number of VoIP users. This is shown in Fig \ref{fig:starvation} with a round robin heuristic, i.e. data flows are given a weight proportional to their time-since-last-scheduled \cite{monikandan2017review, dahlman20185g}. We compare the downlink network throughput and the VoIP user QoS satisfaction rate achieved using QADRA to those achieved using the round-robin and proportional fair heuristics \cite{monikandan2017review}. The proportional fair heuristic is defined as $w_j = \frac{u}{\bar{u}}$, where $u$ and $\bar{u}$ denote the expected throughput and mean throughput, respectively. Finally, in these baselines we also prioritize re-transmissions, however these are rare due to the cellular reception quality.

As defined in Sec. \ref{sec:method}, the reward is measured in terms of the (downlink) throughput of the full buffer traffic, and the QoS satisfaction of VoIP data packets through the two-dimensional vector $\vec{r}=[r_{\text{fb}}, r_{\text{VoIP}}]^T$. We define the QoS requirements of VoIP data flows as a maximum packet delay of \SI{100}{\milli\second}\footnotemark. The reward function for VoIP data packets is therefore given by $r_{\text{VoIP}}(d)=-\Indicator\{d > 0.1\}$, where $d$ is the packet delay measured in seconds. Further, we let $r_{\text{fb}}$ equal the number of bits successfully transferred over the network by the full buffer in the current TTI.

\footnotetext{See 3GPP standard 23.203 "Policy and Charging Control Architecture"}

We let the feature space of data flows consist of the following six variables: (1) time since last scheduled, (2) traffic type (encoded as an integer), (3) number of bits ready for transmission, (4) flag indicating up or downlink flow, (5) flag indicating new transmission, and (6) flag indicating re-transmission. Each data flow is therefore characterized by a feature vector in
\begin{equation}
    \mathcal{R} = [0,\infty)\times\mathbb{Z}\times\mathbb{N}\times\{0,1\}\times\{0,1\}\times\{0,1\}.
\end{equation}

Each experiment was run using four actor processes gathering experiences to the learner process. Every 10th TTI, the network parameters and exploration rate are updated from the learner process. Inspired by \cite{horgan2018distributed}, we vary the exploration rate throughout training. In particular, we let the exploration rate of the $p$th update be given by $\epsilon_p = a^{1+u_p}$, where $u_p = p~\text{mod}~8$. The value of $a$ decreases exponentially from $1.0$ to $0.4$ throughout training. During the first 20000 TTIs of training, we let $\epsilon=1$ in order to build the replay buffer, which has a capacity of 131072 sequences. The training process is only started when the buffer reaches 20000 sequences. Before starting, the mean and standard deviation of the feature variables are computed and then used for normalization of inputs to the networks. Finally, in addition to the DQN, we also implement the double \cite{van2016deep} and prioritized experience replay \cite{schaul2015prioritized} DQN extensions.

The recurrent networks consist of a fully connected feed-forward network with \texttt{ReLU} activations and layer sizes $6\times256\times128$, followed by three gated recurrent units \cite{cho2014learning} with state sizes $64\times32\times32$. The final state is used as the output, i.e. each list is encoded into $\mathbb{R}^{32}$. The fully connected feed-forward Q-Network uses layer sizes $(6+32+32)\times512\times256\times128\times64\times1$ and \texttt{ReLU} activations. The learner process uses an Adam optimizer \cite{kingma2014adam} with a batch size of 32 and learning rate $10^{-4}$. Moreover, we normalize the rewards by factors $200000$ and $0.01$, for the full buffer and VoIP rewards respectively -- however these constants can of course be seen as part of the preference vector, $\omega$.
\subsection{Results}
\begin{figure}
    \centering
    \begin{subfigure}{0.9\linewidth}
        \centering
        \includegraphics[width=\linewidth]{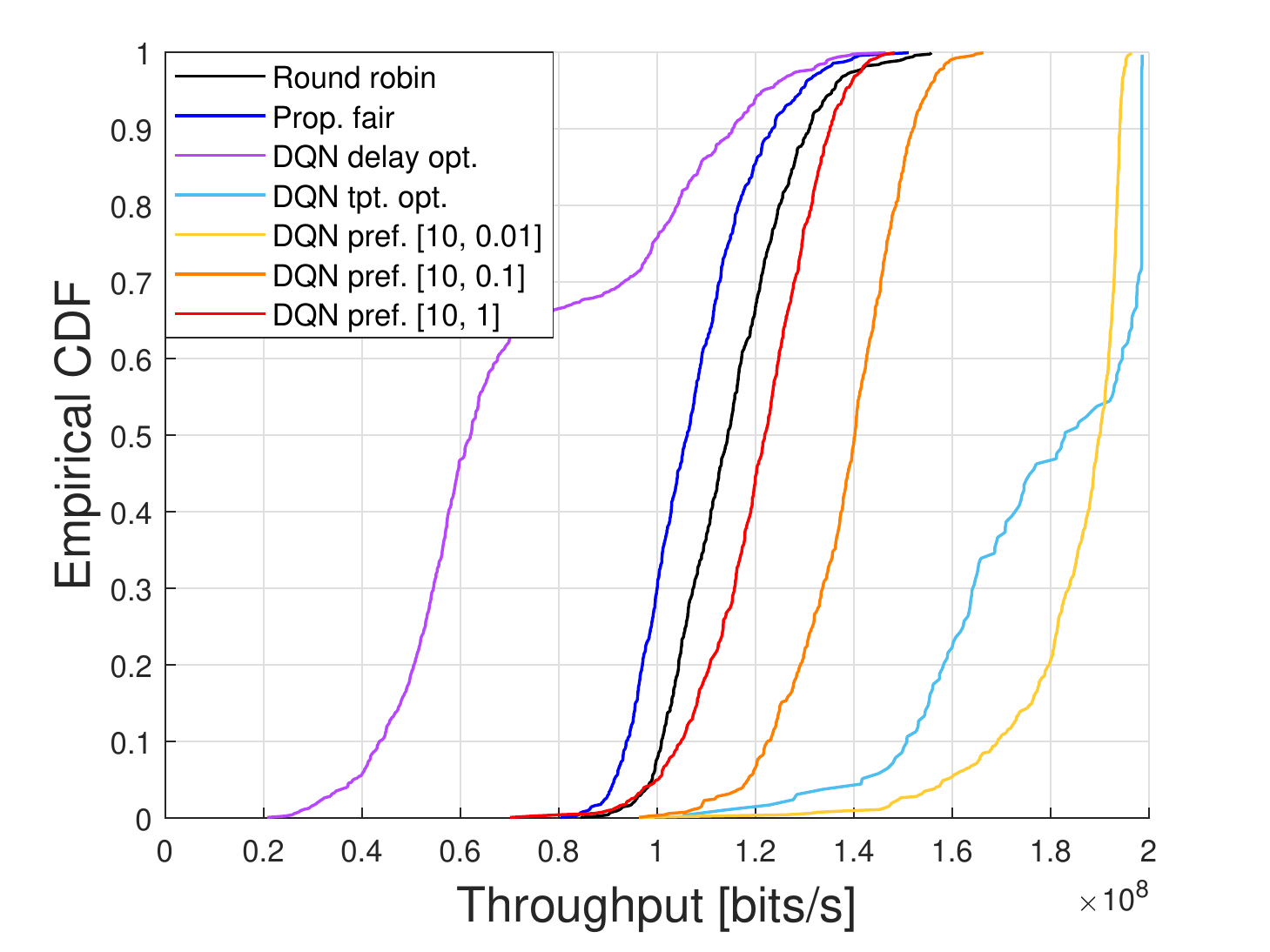}
        \caption{}
        \label{fig:combined:dl}
    \end{subfigure}
    \quad\quad\quad
    \begin{subfigure}{0.9\linewidth}
        \centering
        \includegraphics[width=\linewidth]{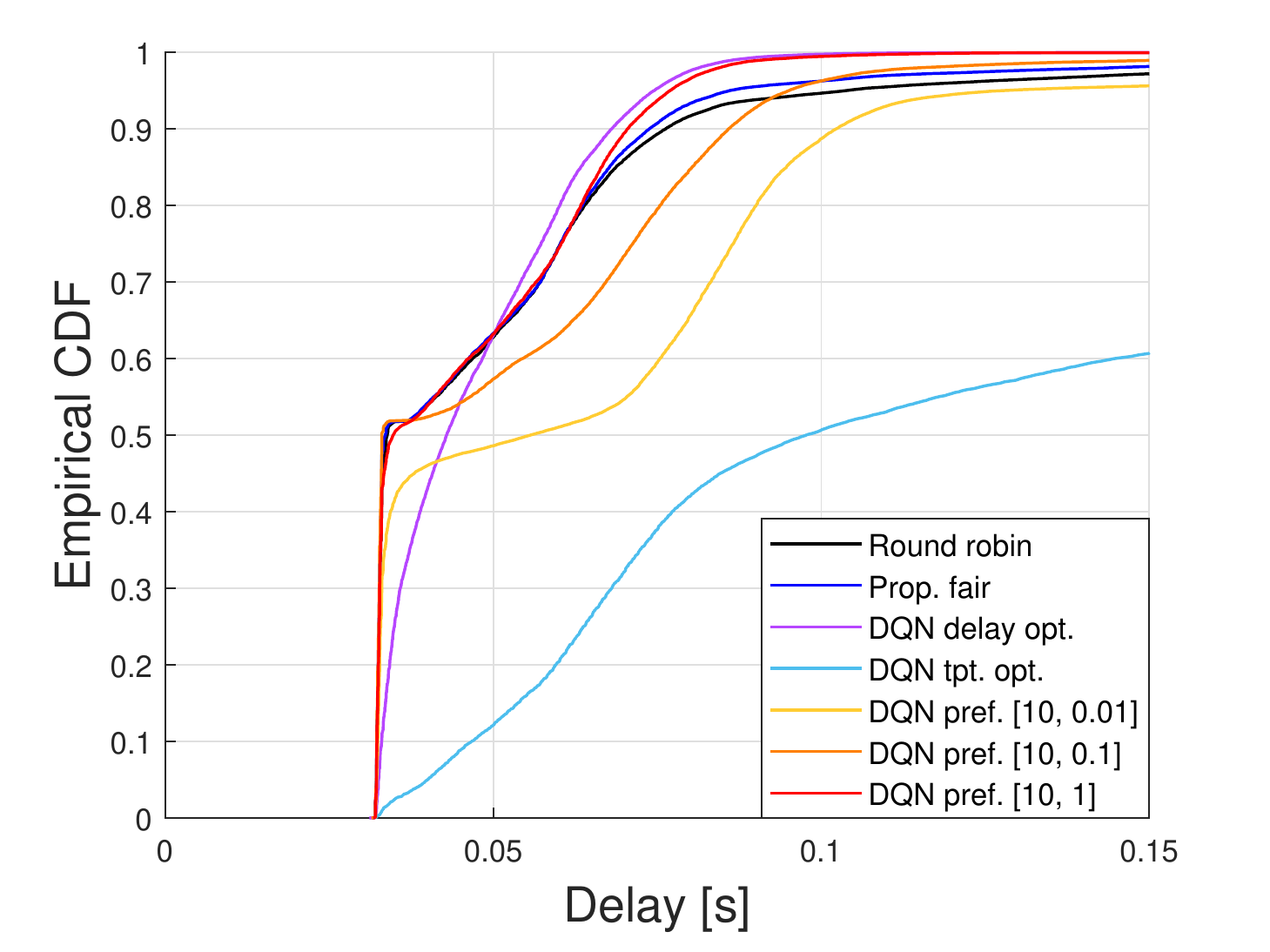}
        \caption{}
        \label{fig:combined:delay}
    \end{subfigure}
    \caption{Empricical cumulative distributions of (a) downlink network throughput and (b) VoIP packet delays for the evaluated techniques.}
    \label{fig:combined}
\end{figure}

Next, we present the numerical results obtained through our experiments. All results are obtained by running the trained agent greedily, i.e. $\epsilon=0$, for 30 simulation seconds. In Fig. \ref{fig:combined}, we present empirical distributions of (\subref{fig:combined:dl}) downlink cell throughput, and (\subref{fig:combined:delay}) VoIP packet delays, when using the evaluated scheduling algorithms. In the context of the QADRA, we first evaluate two extreme configurations: throughput maximization ($\vec{\omega}=[1, 0]^T$) in cyan, and VoIP delay minimization ($\vec{\omega}=[0, 1]^T$) in purple. The results illustrate that the two configurations indeed optimize the throughput and delay respectively, however also results in degraded performance of the non-optimized objective.

We illustrate the performance of three non-trivial preference vectors, $\vec{\omega}^T\in\{[10, 1], [10, 0.1], [10, 0.01]\}$, with yellow, orange, and red respectively. These results are compared to the proportional fair algorithm (dark blue) in Table \ref{tab:results:multi}. In particular, the DRL configuration $\vec{\omega}^T = [10, 0.1]$ results in approximately $30\%$ improved downlink network throughput with no impact (minor improvement) on the QoS satisfaction rate of VoIP data packets. Compared to round robin (black curve), throughput for this configuration is increased by approximately $20\%$ while QoS-affecting delayed packets are reduced by $30\%$. Furthermore, from Table \ref{tab:results:multi} and Fig. \ref{fig:combined}, we observe that tuning the preference vector provides fine-grained control over the scheduler QoS performance. Specifically, an increased weight on VoIP improves QoS performance at the cost of downlink throughput and vice versa.

\begin{table}
    \centering
    \resizebox{0.47\textwidth}{!}{\begin{tabular}{c|c|c|c|c}
        Preference vector, $\vec{\omega}^T$ & tpt 10\% & tpt 50\% & tpt 90\% & Delay \\ \hline
        $[10, 1]$ & $+12\%$ & $+14.9\%$ & $+9.6\%$ & $-85.4\%$ \\
        $[10, 0.1]$ & $+30.6\%$ & $+32.0\%$ & $+23.2\%$ & $-0.4\%$ \\
        $[10, 0.01]$ & $+80.6\%$ & $+72.4\%$ & $+57.4\%$ & $+199.3\%$ 
    \end{tabular}}
    \caption{Relative change in downlink throughput (tpt) percentiles and fraction of VoIP packets delayed longer than \SI{100}{\milli\second}, compared to proportional fair scheduling.}
    \label{tab:results:multi}
\end{table}

\section{Conclusions} \label{sec:conclusion}
In this paper, we have proposed a new approach to time domain scheduling in cellular networks using deep reinforcement learning. In contrast to current state-of-the-art algorithms, sorting data flows implicitly through heuristics, our proposed algorithm \textit{explicitly} sorts data flows through a selection sort-like process. Moreover, we train our algorithm end-to-end as to explicitly optimize user experience and network performance, thus limiting the need for human interference and/or manual tuning of the implemented scheduler. Finally, we allow a network operator to control fundamental network performance trade-offs, explicitly and reliably, through the preference vector.

We numerically evaluate our scheduling algorithm in a full scale, near-real, NR system level simulator, and compare it to two state-of-the-art scheduling algorithms. Specifically, we consider a network of ten VoIP users and one full buffer downlink flow. The results demonstrate that our scheduling approach provides a significant boost in network performance. This is illustrated with an increased cell throughput of approximately $30\%$ while simultaneously keeping QoS of VoIP data flows unaffected (see Table \ref{tab:results:multi}).

\subsection{Future work}
We consider two directions for future work. Firstly, although evaluated in a rather specific resource starvation scenario, we believe that our algorithm is capable of learning optimal scheduling in more complex network scenarios as well, involving increased number of data flow types and users. Secondly, extending the algorithm to consider varying optimization objectives may allow for varying preference vectors at inference time, thus more control of fundamental trade-offs. Incorporating the work of Yang \& Sun \cite{yang2019generalized} into the algorithm may yield interesting results.

\FloatBarrier

\printbibliography

@article{monikandan2017review,
  title={A review of {MAC} scheduling algorithms in {LTE} system},
  author={Monikandan, B and Sivasubramaznian, A and Babu, SPK and others},
  journal={International Journal on Advanced Science, Engineering and Information Technology},
  volume={7},
  number={3},
  pages={1056--1068},
  year={2017}
}

@book{dahlman20185g,
  title={5G NR: The next generation wireless access technology},
  author={Dahlman, Erik and Parkvall, Stefan and Skold, Johan},
  year={2018},
  publisher={Academic Press}
}

@article{mnih2013playing,
  title={Playing atari with deep reinforcement learning},
  author={Mnih, Volodymyr and Kavukcuoglu, Koray and Silver, David and Graves, Alex and Antonoglou, Ioannis and Wierstra, Daan and Riedmiller, Martin},
  journal={arXiv preprint arXiv:1312.5602},
  year={2013}
}

@inproceedings{van2016deep,
  title={Deep reinforcement learning with double {Q-learning}},
  author={Van Hasselt, Hado and Guez, Arthur and Silver, David},
  booktitle={Thirtieth AAAI conference on artificial intelligence},
  year={2016}
}

@article{schaul2015prioritized,
  title={Prioritized experience replay},
  author={Schaul, Tom and Quan, John and Antonoglou, Ioannis and Silver, David},
  journal={arXiv preprint arXiv:1511.05952},
  year={2015}
}

@article{capozzi2012downlink,
  title={Downlink packet scheduling in {LTE} cellular networks: Key design issues and a survey},
  author={Capozzi, Francesco and Piro, Giuseppe and Grieco, Luigi Alfredo and Boggia, Gennaro and Camarda, Pietro},
  journal={IEEE communications surveys \& tutorials},
  volume={15},
  number={2},
  pages={678--700},
  year={2012},
  publisher={IEEE}
}

@article{grondalen2016scheduling,
  title={Scheduling policies in time and frequency domains for {LTE} downlink channel: a performance comparison},
  author={Gr{\o}ndalen, Ole and Zanella, Andrea and Mahmood, Kashif and Carpin, Mattia and Rasool, Jawad and {\O}sterb{\o}, Olav N},
  journal={IEEE Transactions on Vehicular Technology},
  volume={66},
  number={4},
  pages={3345--3360},
  year={2016},
  publisher={IEEE}
}

@article{cho2014learning,
  title={Learning phrase representations using {RNN} encoder-decoder for statistical machine translation},
  author={Cho, Kyunghyun and Van Merri{\"e}nboer, Bart and Gulcehre, Caglar and Bahdanau, Dzmitry and Bougares, Fethi and Schwenk, Holger and Bengio, Yoshua},
  journal={arXiv preprint arXiv:1406.1078},
  year={2014}
}

@article{horgan2018distributed,
  title={Distributed prioritized experience replay},
  author={Horgan, Dan and Quan, John and Budden, David and Barth-Maron, Gabriel and Hessel, Matteo and Van Hasselt, Hado and Silver, David},
  journal={arXiv preprint arXiv:1803.00933},
  year={2018}
}

@book{sutton2018reinforcement,
  title={Reinforcement Learning: An Introduction},
  author={Sutton, R.S. and Barto, A.G.},
  isbn={9780262039246},
  lccn={2018023826},
  year={2018},
  publisher={MIT Press}
}

@article{kingma2014adam,
  title={{ADAM}: A method for stochastic optimization},
  author={Kingma, Diederik P and Ba, Jimmy},
  journal={arXiv preprint arXiv:1412.6980},
  year={2014}
}

@inproceedings{yang2019generalized,
  title={A Generalized Algorithm for Multi-Objective Reinforcement Learning and Policy Adaptation},
  author={Yang, Runzhe and Sun, Xingyuan and Narasimhan, Karthik},
  booktitle={Advances in Neural Information Processing Systems},
  pages={14610--14621},
  year={2019}
}

@article{watkins1989learning,
  title={Learning from delayed rewards},
  author={Watkins, Christopher John Cornish Hellaby},
  year={1989},
  publisher={King's College, Cambridge}
}

@article{pedroso2019low,
  title={A Low-Complexity Scheduler to Improve the Number of Satisfied Video Streaming Users in LTE},
  author={Pedroso, Carlos M and da Silva, Carlos A Gouvea and Junior, Joel A Barbosa and Mafra, Samuel B},
  journal={Wireless Personal Communications},
  volume={109},
  number={2},
  pages={1121--1132},
  year={2019},
  publisher={Springer}
}

\end{document}